\documentstyle[aps,epsfig,a4]{revtex}
\begin{document}
\draft

\title{Analysis of the freeze-out parameters for RHIC, SPS and AGS based on
${ {dE_{T}} \over {d\eta} } / { {dN_{ch}} \over {d\eta} }$ ratio
measurements }
\author{Dariusz Prorok}
\address{Institute of Theoretical Physics, University of
Wroc{\l}aw,\\ Pl.Maksa Borna 9, 50-204  Wroc{\l}aw, Poland}
\date{May 16, 2003}
\maketitle
\begin{abstract}
The ratio ${ {dE_{T}} \over {d\eta} } / { {dN_{ch}} \over {d\eta}
}$ is analyzed in the framework of a single-freeze-out thermal
hadron gas model. Decays of hadron resonances are taken into
account in evaluations of this ratio. The predictions of the model
at the freeze-out parameters, established previously from observed
particle yields, agree very well with the ratio measured at RHIC,
SPS and AGS.
\end{abstract}


In this paper, a statistical model is tested in recovering values
of the ratio ${ {dE_{T}} \over {d\eta} }_{\mid mid} / { {dN_{ch}}
\over {d\eta} }_{\mid mid}$ measured at RHIC \cite{Adcox:2001ry},
SPS \cite{Aggarwal:2000bc} and AGS \cite{Barrette:1994kr} ({\it
mid} means the midrapidity region). For RHIC and SPS the ratio
equals about $0.8$ GeV, but for AGS it is $0.72$ GeV. So far, the
model has been applied successfully in explanation of particle
ratios and distributions observed in heavy-ion collisions
\cite{Braun-Munzinger:1994xr,Cleymans:1996cd,Braun-Munzinger:1999qy,Braun-Munzinger:2001ip,Florkowski:2001fp,Michalec:2001um,Broniowski:2001we,Broniowski:2001uk}.
Since the transverse energy measurement is independent and easier
(no particle identification is necessary), it gives an unique
opportunity to verify the concept of the appearance of a thermal
system during such collisions.

The statistical model with single freeze-out is used
\cite{Florkowski:2001fp,Michalec:2001um,Broniowski:2001we,Broniowski:2001uk}.
The model reproduces very well ratios and $p_{T}$ spectra of
particles observed at RHIC
\cite{Florkowski:2001fp,Broniowski:2001we,Broniowski:2001uk}. The
main assumption of the model is the simultaneous occurrence of
chemical and thermal freeze-outs. The new data on $K^{*}(892)^{0}$
production revealed by the STAR Collaboration \cite{Adler:2002sw}
support strongly this assumption. Since ${ {dE_{T}} \over {d\eta}
}_{\mid mid} / { {dN_{ch}} \over {d\eta} }_{\mid mid}$
measurements have been done at midrapidity, the presented analysis
is valid in the Central Rapidity Region (CRR) of considered
collisions.

Therefore, it is assumed that a noninteracting gas of stable
hadrons and resonances at chemical and thermal equilibrium is
present at the CRR. For consistency with previous works
\cite{Braun-Munzinger:1994xr,Cleymans:1996cd,Braun-Munzinger:1999qy,Braun-Munzinger:2001ip,Florkowski:2001fp,Michalec:2001um},
where freeze-out parameters were found out from particle ratio
measurements for a static fireball, this is also the case
considered here. Then the distributions of various species of
primordial particles are given by the usual ideal-gas formulae.
Only baryon number $\mu_{B}$ and strangeness $\mu_{S}$ chemical
potentials are taken into account here. The isospin chemical
potential $\mu_{I_{3}}$ has very low value in analyzed cases
\cite{Braun-Munzinger:1999qy,Florkowski:2001fp} and therefore can
be neglected. For given $T$ and $\mu_{B}$, $\mu_{S}$ is determined
from the requirement that the overall strangeness of the gas
equals zero. In this way, the temperature $T$ and the baryon
chemical potential $\mu_{B}$ are the only independent parameters
of the model.

Theoretically, the transverse energy is defined as the sum of
transverse masses of all $L$ interacting and produced particles
\cite{Albrecht:1991fg},

\begin{equation}
E_{T}^{th} = \sum_{i = 1}^{L} \sqrt{m_{i}^{2}+(p_{T}^{i})^{2}} \;,
\label{Thent}
\end{equation}

\noindent where $m_{i}$ and $p_{T}^{i}$ are the mass and
transverse momentum, respectively. But experimentally, the
measured quantity is

\begin{equation}
E_{T} = \sum_{i = 1}^{L} E_{i} \cdot \sin{\theta_{i}} \;,
\label{Exent}
\end{equation}

\noindent where $\theta_{i}$ is the polar angel and $E_{i}$
denotes the total energy. Precisely, what is measured in a
calorimeter is the kinetic energy for nucleons and the total
energy for all other particles \cite{Adcox:2001ry}.

If one deals with the thermal system, definitions (\ref{Thent})
and (\ref{Exent}) should be generalized appropriately. The
transverse energy density $\epsilon_{T}^{i}$ of the particles of
species $i$ can be defined at the temperature $T$ and the baryon
chemical potential $\mu_{B}$ as ($\hbar=c=1$ always)

\begin{equation}
\epsilon_{T}^{i}= (2s_{i}+1) \int d^{3}\vec{p}\;
e_{T}^{i}(\vec{p})\; f_{i}(p;T,\mu_{B})\;, \label{energyt}
\end{equation}

\noindent where $e_{T}^{i}(\vec{p})$ is the suitable expression
for the transverse energy of the particle and $m_{i}$, $s_{i}$ and
$f_{i}(p;T,\mu_{B})$ are its mass, spin and momentum distribution
(which is isotropic here), respectively. Since the theoretical
estimates should correspond to what is actually measured,
$e_{T}^{i}$ takes the form

\begin{equation}
e_{T}^{i}(\vec{p}) = E_{i} \cdot \sin{\theta} = E_{i} \cdot
{{p_{T}} \over p} \; \label{trueet}
\end{equation}

\noindent in the analyzed model.

The thermal system ceases at the freeze-out and there are only
free escaping particles instead of the fireball. The measured
${{dE_{T}} \over {d\eta} }_{\mid mid}$ is fed from two sources:
(a) stable hadrons which survived until catching in a detector,
(b) secondaries produced by decays and sequential decays of
primordial resonances after the freeze-out. Therefore, if the
contribution to the transverse energy from particles (a) is
described, the distribution $f_{i}$ in (\ref{energyt}) is either a
Bose-Einstein or a Fermi-Dirac distribution at the freeze-out. But
if the contribution from particles (b) is considered, the
distribution $f_{i}$ is the spectrum of the finally detected
secondaries and can be obtained from the elementary kinematics of
a many-body decay or from the superposition of two or more such
decays (for details, see
\cite{Florkowski:2001fp,Broniowski:2001uk}; also
\cite{Sollfrank:1990qz,DeWit:it} can be very useful). In fact, if
one considers detected species $i$, then $f_{i}$ is the sum of
final $i$'s spectra resulting from a single decay or a cascade.
The sum is taken over all such decays and cascades of resonances
in which at least one of the final secondaries is of the $i$ kind.

Since ${ {dN_{ch}} \over {d\eta} }_{\mid mid}$ has also its origin
in the above-mentioned sources (a) and (b), to define properly the
density of charged particles, decays should be also taken into
account. Thus the density of the measured charged particles of
species $j$ reads

\begin{equation}
n^{j} = n_{primordial}^{j} + \sum_{i} \alpha(j,i)\;
n_{primordial}^{i} \;, \label{nchj}
\end{equation}

\noindent where $n_{primordial}^{i}$ is the density of the $i$th
particle species at the freeze-out and $\alpha(j,i)$ is the final
number of particles of species $j$ which can be received from all
possible simple or sequential decays of particle $i$. The density
$n_{primordial}^{i}$ is given by the usual integral of either a
Bose-Einstein or a Fermi-Dirac distribution. Now, in the
midrapidity region, the theoretical equivalent of ${ {dE_{T}}
\over {d\eta} }_{\mid mid} / { {dN_{ch}} \over {d\eta} }_{\mid
mid}$ can be postulated as

\begin{equation}
{{{ {dE_{T}} \over {d\eta} }_{\mid mid}} \over {{ {dN_{ch}} \over
{d\eta} }_{\mid mid}}} \equiv { {\epsilon_{T}} \over {n_{ch}} }
\;, \label{thratio}
\end{equation}

\noindent where the transverse energy density $\epsilon_{T}$ and
the density of charged particles $n_{ch}$ are given by the
expressions

\begin{equation}
\epsilon_{T} = \sum_{i \in A} \epsilon_{T}^{i} \;, \label{enertot}
\end{equation}

\begin{equation}
n_{ch} = \sum_{j \in B} n^{j} \;. \label{nchtot}
\end{equation}

\noindent Note that there are two different sets of final
particles, $A$ and $B$ ($B \subset A$). $B$ denotes final charged
particles and these are $\pi^{+},\; \pi^{-},\; K^{+},\; K^{-},\;
p$ and $\bar{p}\;$, whereas $A$ also includes photons,
$K_{L}^{0},\; n$ and $\bar{n}\;$ \cite{Adcox:2001ry}.

To proceed further, some simplifications are necessary. This is
because the complete treatment of resonance decays in
$\epsilon_{T}$ is complex and therefore it consumes a lot of
computer working time in numerical calculations. So the initial
set of resonances should be as small as possible. The lifetime of
at least $10$ fm is chosen as the necessary condition, with the
exception of $K^{*}(892)$ mesons, since one of them is measured at
RHIC \cite{Adler:2002sw}. This reduces constituents of the hadron
gas to $40$ species, but note that all particles listed in
\cite{Braun-Munzinger:1994xr,Cleymans:1996cd,Braun-Munzinger:1999qy,Braun-Munzinger:2001ip,Florkowski:2001fp}
are included. Of course, this condition is arbitrary but it makes
sense because most neglected resonances have the lifetime of the
order of a few fm, so one may assume they decay already at the
pre-equilibrium stage. Anyway, on the basis of the analysis with
decays not included in $\epsilon_{T}$, it has been found that the
ratio (\ref{thratio}) changes slowly with the number of species of
the hadron gas in the considered region of $T$ and $\mu_{B}$.

In the following, a point-like gas is assumed. It has been checked
that for the excluded volume hadron gas model
\cite{Hagedorn:1978kc,Hagedorn:1980kb,Yen:1997rv} the results are
exactly the same. There are two reasons for that: the first, the
volume corrections placed in denominators of expressions for
various densities cancel with each other in a ratio; the second,
the eigenvolume of a hadron and the pressure in the considered
region of $T$ and $\mu_{B}$ are so small that their product
correction to the chemical potential is negligible.

As it has been already mentioned, the main difficulty in complete
treatment of decays in numerical evaluations of $\epsilon_{T}$ is
their complexity. Therefore some further simplifications should be
done. First of all, some decays and cascades are neglected: (i)
four-body decays, (ii) superpositions of two three-body decays,
(iii) superpositions of two three-body and one two-body decays,
(iv) superpositions of four two-body decays, (v) some decays of
heavy resonances with very small branching ratios. Their maximal
contribution to $\epsilon_{T}$ has been evaluated as $\leq 2 \%$.
Thus, the real $\epsilon_{T}$ can be at most $2 \%$ higher than
its evaluation. It should be stressed that the above-mentioned
simplifications are done only in calculations of $\epsilon_{T}$,
whereas $n_{ch}$ given by (\ref{nchj}) and (\ref{nchtot}) includes
all possible decays and cascades. This means that the presented
values of the ratio (\ref{thratio}) (and (\ref{booratio})) can be
also at most $2 \%$ higher.

In application to heavy-ion collisions, it is assumed that the
rest frame of the hadron gas is the c.m.s of two colliding ions.
But RHIC is the opposite beam experiment, whereas SPS and AGS are
the fixed target ones. So, the laboratory frame is the c.m.s only
in the RHIC case. Since the measurement is done in the laboratory
frame, to treat SPS and AGS cases properly, it is assumed that
there is the overall uniform flow (of the gas) with constant
velocity $v$ equal to the velocity $v_{c.m.s}$ of the c.m.s
relative to the target. The $v_{c.m.s}$ is calculated for $158
\cdot A$ GeV Pb-Pb collisions at SPS (this results in $v_{c.m.s} =
0.994$) and for $11 \cdot A$ GeV Au-Au collisions at AGS
($v_{c.m.s} = 0.918$).

Now one applies the general description founded in
\cite{Cooper:1974mv} and developed in \cite{Broniowski:2002nf} for
the case with decays taken into account. The invariant
distribution of the measured particles of species $j$ has the form
\cite{Broniowski:2002nf}

\begin{equation}
E_{j}{ {dN_{j}} \over {d^{3}\vec{p}} }=\int
p^{\mu}d\sigma_{\mu}\;f_{j}(p \cdot u) \;, \label{Cooper}
\end{equation}

\noindent where $d\sigma_{\mu}$ is the normal vector on a
freeze-out hypersurface, $p \cdot u = p^{\mu}u_{\mu}$ , $u$ is the
appropriate four-velocity and $f_{j}$ is the final momentum
distribution of the particle in question. The final means here
that $f_{j}$ is the sum of primordial and decay contributions to
the particle distribution. For the static (homogeneous) fireball,
the freeze-out hypersurface is simply a volume $V^{*}$ at a
freeze-out time $t_{f.o.}^{*}$ (stars denote quantities in the
comoving frame, which is the local rest frame of the gas and also
the c.m.s here). The spatial coordinates
$x^{*1},\;x^{*2},\;x^{*3}$ are the parameters of the hypersurface
in this case. Thus the normal vector reads
$d\sigma_{\mu}^{*}=(d^{3}\vec{x}^{*},0,0,0)$. In the laboratory
frame $d\sigma_{\mu}=(\gamma d^{3}\vec{x}^{*},0,0,-\gamma v
d^{3}\vec{x}^{*})$, $\gamma = (1 - v^{2})^{-1/2}$, and since
$u_{\mu} = \gamma (1,0,0,-v)$ in the considered case, the normal
vector is proportional to the four-velocity,
$d\sigma_{\mu}=u_{\mu}d^{3}\vec{x}^{*}$. This is the necessary
condition for the invariant distribution expression of particles
$j$ to have the form (\ref{Cooper}), where $f_{j}$ is calculated
in the local rest frame of the gas (for more details, see
\cite{Broniowski:2002nf}). Note also that since $u^{\mu}u_{\mu} =
1$, the normal vector is time-like, so the hypersurface consists
only of a space-like part. Thus the conceptual problems with
time-like parts of a hypersurface are avoided (for discussion of
the subject, see e.g. \cite{Rischke:1996em,Anderlik:1998et} and
references therein). Now the invariant distribution of the
particles of species $j$ reads

\begin{equation}
E_{j}{ {dN_{j}} \over {d^{3}\vec{p}} }=\int_{V^{*}}
d^{3}\vec{x}^{*} (p \cdot u)\;f_{j}(p \cdot u)=V^{*} (p \cdot
u)\;f_{j}(p \cdot u) \;, \label{invdistr}
\end{equation}

\noindent where $p \cdot u=\gamma \cdot (E_{j} - p_{z}
v)=E_{j}^{*}$ is the energy in the comoving frame. From
(\ref{invdistr}) the multiplicity of the $j$-th particles can be
obtained as

\begin{equation}
N_{j}=V^{*} \int { {d^{3}\vec{p}} \over {E_{j}}}(p \cdot
u)\;f_{j}(p \cdot u) = V^{*} \int
d^{3}\vec{p}^{*}\;f_{j}(E_{j}^{*}) = V^{*} \cdot n^{j}\;,
\label{multipli}
\end{equation}

\noindent where the last equality holds because distributions of
particles depend here only on the magnitude of the three-momentum,
so $f_{j}(E_{j}^{*})=f_{j}(p^{*})$. Similarly, the final
transverse energy carried by the particles of species $j$ is
calculated as

\begin{equation}
E_{T}^{j}=V^{*} \int { {d^{3}\vec{p}} \over {E_{j}}}\;E_{j}
{{p_{T}} \over p}(p \cdot u)\;f_{j}(p \cdot u) = V^{*} \int
d^{3}\vec{p}^{*}\;E_{j} {{p_{T}} \over p} f_{j}(p^{*})\;,
\label{tren}
\end{equation}

\noindent where $E_{j}=\gamma \cdot (E_{j}^{*} + p_{z}^{*} v)$,
$p_{T}=p_{T}^{*}$ and $p=\sqrt{E_{j}^{2}-m_{j}^{2}}$. And finally,
the transverse energy, the multiplicity of charged particles and
the ratio (\ref{thratio}) are given by the expressions

\begin{equation}
E_{T} = \sum_{j \in A} E_{T}^{j} \;, \label{tottren}
\end{equation}

\begin{equation}
N_{ch} = \sum_{j \in B} N_{j} \;, \label{totchar}
\end{equation}

\begin{equation}
{{{ {dE_{T}} \over {d\eta} }_{\mid mid}} \over {{ {dN_{ch}} \over
{d\eta} }_{\mid mid}}} \equiv { {E_{T}} \over {N_{ch}} } \;,
\label{booratio}
\end{equation}

\noindent respectively. Note that the dependence on the volume
$V^{*}$ has disappeared in the ratio (\ref{booratio}) and for
$v=0$ the formula (\ref{thratio}) has been recovered.

\vskip 1cm

\begin{center}
\begin{tabular}{|c|c|c|c|} \hline { \hbox to 3truecm{}}& \multicolumn{2}{c|}{} &
 \cr & \multicolumn{2}{c|}{$E_{T}/N_{ch}$ [GeV]} &
 \cr & \multicolumn{2}{c|}{($v_{c.m.s}$ appropriate) } & ${ {dE_{T}} \over {d\eta}} / {{dN_{ch}} \over {d\eta} }$
  \cr \cline{2-3} & { \hbox to 2truecm{}} & $E \rightarrow E-m_{n}$ &
 \cr & & for nucleons & [GeV]
\cr \hline $T = 175$ MeV & & & \cr $\mu_{B} = 51$ MeV & 0.88 &
0.80 & $0.8^{+0.08}_{-0.06}$ \protect\cite{Adcox:2001ry} \cr RHIC
\protect\cite{Braun-Munzinger:2001ip} & & & \cr \hline $T = 165$
MeV & & & \cr $\mu_{B} = 41$ MeV & 0.84 & 0.77 &
$0.8^{+0.08}_{-0.06}$ \protect\cite{Adcox:2001ry} \cr RHIC
\protect\cite{Florkowski:2001fp} & & & \cr \hline $T = 168$ MeV &
& & \cr $\mu_{B} = 266$ MeV & 0.76 & 0.75 & $0.8 \pm 0.2$
\protect\cite{Aggarwal:2000bc} \cr SPS
\protect\cite{Braun-Munzinger:1999qy} & & & \cr \hline $T = 164$
MeV & & & \cr $\mu_{B} = 234$ MeV & 0.74 & 0.73 & $0.8 \pm 0.2$
\protect\cite{Aggarwal:2000bc} \cr SPS
\protect\cite{Michalec:2001um} & & & \cr \hline $T = 130$ MeV & &
& \cr $\mu_{B} = 540$ MeV & 0.81 & 0.66 & $0.72 \pm 0.08$
\protect\cite{Barrette:1994kr} \cr AGS
\protect\cite{Braun-Munzinger:1994xr} & & & \cr \hline $T = 110$
MeV & & & \cr $\mu_{B} = 540$ MeV & 0.70 & 0.57 & $0.72 \pm 0.08$
\protect\cite{Barrette:1994kr} \cr AGS
\protect\cite{Cleymans:1996cd} & & & \cr \hline
\end{tabular}
\end{center}
\begin{table}
\caption{Values of ${{E_{T}} \over {N_{ch}}}$ calculated with the
use of the formula (\ref{trueet}) (second column) and the
realistic version of the formula (\ref{trueet}) with the kinetic
energy instead of the total energy for nucleons (third column). In
the first column estimates of freeze-out parameters obtained from
the analysis of particle ratios
\protect\cite{Braun-Munzinger:1994xr,Cleymans:1996cd,Braun-Munzinger:1999qy,Braun-Munzinger:2001ip,Florkowski:2001fp,Michalec:2001um}
are listed. In the last column experimental data are given. The
velocity $v_{c.m.s}$ is the velocity of the center of mass of
colliding nuclei with respect to the laboratory frame and equals
respectively: 0 for RHIC, 0.994 for SPS and 0.918 for AGS.}
\label{Table1}
\end{table}

\begin{figure}
\begin{center}{
{\epsfig{file=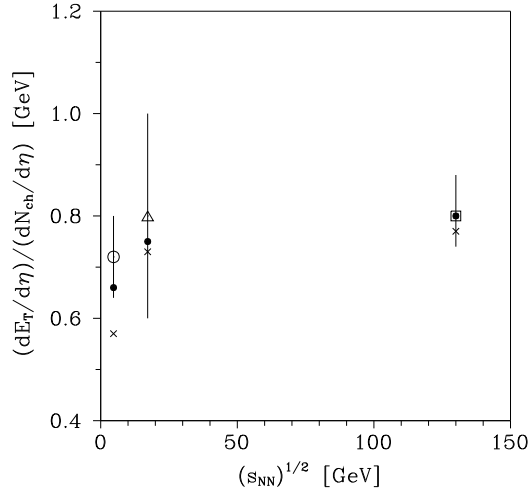,width=7cm}} }\end{center}
\caption{Values of $E_{T} / N_{ch}$ from the third column of
TABLE\,\ref{Table1}. Black dots and crosses denote evaluations of
the ratio at higher and lower temperature for a given collider,
respectively (see the first column of TABLE\,\ref{Table1}). Also
data points for AGS \protect\cite{Barrette:1994kr} (circle), SPS
\protect\cite{Aggarwal:2000bc} (triangle) and RHIC
\protect\cite{Adcox:2001ry} (square) are depicted.}
\label{Fig.1.}
\end{figure}

The final results of calculations following from the formula
(\ref{booratio}) are presented in TABLE\,\ref{Table1}. In the
first column estimates of freeze-out parameters obtained from the
analysis of particle ratios
\cite{Braun-Munzinger:1994xr,Cleymans:1996cd,Braun-Munzinger:1999qy,Braun-Munzinger:2001ip,Florkowski:2001fp,Michalec:2001um}
are listed. In the second column corresponding values of ${{E_{T}}
\over {N_{ch}}}$ calculated with the use of (\ref{trueet}) are
placed. In the third column of TABLE\,\ref{Table1}, the most
realistic results are presented, namely the fact that for nucleons
only the kinetic energy is measured \cite{Adcox:2001ry} is taken
into account. The appropriate $v_{c.m.s}$ is put for each
collider. For clearness, these results have been also depicted
together with the data in Fig.\,\ref{Fig.1.}.

Generally, very good agreement of the obtained predictions with
the data has been reached. One can notice that the accuracy of
theoretical evaluations rises with the collision energy. For AGS,
the ratio $E_{T} / N_{ch}$ is within error bars of the
experimental point only for the higher temperature freeze-out,
whereas for RHIC both estimates fit very well. This could mean
that AGS energy is the low limit of applicability of a statistical
model. Alternatively, one could call the application of the same
gas for both RHIC and AGS in question. This is because of a
different baryon content of each case. The AGS gas should be much
more rich with baryons than the RHIC one. If one considers
contributions to primordial $E_{T}$ (i.e. decays are not included
in calculation of $E_{T}$) from constituents of the gas assumed
here, nucleons weight most for AGS, but pions are the biggest
fraction in the RHIC case. So, probably additional baryon
resonances should be included in a gas in the AGS case. In fact,
some preliminary estimates of primordial $E_{T}$ indicate that by
adding more species into the gas one could increase the $E_{T} /
N_{ch}$ ratio at the AGS freeze-out. Also if one takes the
expansion of the gas into account one might improve the results
for the AGS case. Roughly speaking, the expansion produces
additional energy, so it could increase $E_{T}$. These problems
need much more detailed analysis and will be under further
investigation.

To conclude, a statistical model has been used to reproduce the
ratio ${ {dE_{T}} \over {d\eta} } / { {dN_{ch}} \over {d\eta} }$
measured at RHIC, SPS and AGS. The importance of presented
analysis lies in the fact that the ratio is an independent
observable, so it can be used as a new tool to verify the
consistence of predictions of a statistical model for all
colliders simultaneously. The point-like non-interacting hadron
gas with $40$ species has been used in final calculations. Decays
and sequential decays of constituents of the gas have been taken
into account. In spite of the simplicity of the model, theoretical
predictions for $E_{T} / N_{ch}$ agree very well with the data for
the wide range of collision energies starting from AGS up to RHIC.
The predictions have been made at the previous estimates of
freeze-out parameters obtained from the analysis of measured
particle ratios for RHIC
\cite{Braun-Munzinger:2001ip,Florkowski:2001fp}, SPS
\cite{Braun-Munzinger:1999qy,Michalec:2001um} and AGS
\cite{Braun-Munzinger:1994xr,Cleymans:1996cd}. This means that the
applicability of a statistical model to heavy-ion collisions has
been confirmed strongly in an independent way.

The author acknowledges very stimulating discussions with Peter
Braun-Munzinger and Ludwik Turko. This work was supported in part
by the Polish Committee for Scientific Research under Contract No.
KBN - 2 P03B 030 18.

\end{document}